\documentclass[12pt]{article}
\usepackage{makeidx}

\title{Spontaneous Symmetry Breaking in Quantum Systems. A review for Scholarpedia} 
\author{F. Strocchi \\ INFN, Sezione di Pisa, Pisa Italy}

\makeglossary
\date{}
\newtheorem{Theorem}{Theorem}[section]

\usepackage{latexsym}
\usepackage[T1]{fontenc}
\usepackage{amsmath}
\usepackage{amssymb}

\def \AO {{\cal A}({\cal O})}
\def \AO' {{\cal A}({\cal O}')}

\def \limR {\lim_{R \ra \infty}}

\def \be {\begin{equation}}
\def \ee {\end{equation}}

\def \ra {\rightarrow}

\def \eqq {\equiv}

\def \a {{\alpha}}
\def \b {{\beta}}

\def \d {{\delta}}
\def \eps {{\varepsilon}}

\def \l {{\lambda}}

\def \ph {{\varphi}}

\def \A {{\cal A}}
\def \B {{\cal B}}

\def \D {{\cal D}}
\def \F {{\cal F}}

\def \H {\mbox{${\cal H}$}}

\def \O {{\cal O}}

\def \S {{\cal S}}

\def \id {{\bf 1 }}

\def \Psio {{\Psi_0}}

\def \d^nu {{\partial^\nu}}
\def \d^la {{\partial^\lambda}}
\def \d^o {{\partial^0}}

\def \Fmn {{F_{\mu\,\nu}}}

\def \abf {{\bf a}}

\def \k {{\bf k}}

\def \j   {{\bf j}}

\def \x {{\bf x}}
\def \y {{\bf y}}

\def \Cbf {{\bf C}}

\def \Rbf {{\bf R}}

\def\doppio#1{{\rm I}\kern-.1667em{\rm #1}}

\def\Q{\text{Q}\kern-.52em
    \text{\vrule height1.5ex width.5pt depth0pt}\kern.45em}

\def\dZ{{\mathchoice {\hbox{$\Ss\textstyle Z\kern-0.4em Z$}}
{\hbox{$\Ss\textstyle Z\kern-0.4em Z$}} {\hbox{$\Ss\scriptstyle
Z\kern-0.25em Z$}} {\hbox{$\Ss\scriptscriptstyle Z\kern-0.2em
Z$}}}}

\def\dC{{\mathchoice{\hbox{$\rm\textstyle\text{\kern.35em\vrule
   height1.5ex width.5pt depth0pt\kern-.35em C}$}}
{\hbox{$\rm\textstyle\text{\kern.35em\vrule
   height1.5ex width.5pt depth0pt\kern-.35em C}$}}
{\hbox{$\rm\scriptstyle\text{\kern.35em\vrule
   height1.5ex width.3pt depth0pt\kern-.35em C}$}}
{\hbox{$\rm\scriptscriptstyle\text{\kern.35em\vrule
   height1.5ex width.2pt depth0pt\kern-.35em C}$}}}}

\makeatletter

\@addtoreset{equation}{section}

\begin{document}

\maketitle
\begin{abstract}
The mechanism of spontaneous symmetry breaking in quantum systems is briefly reviewed, rectifying  part of the standard wisdom on  logical and mathematical grounds. The crucial role of the localization properties of the time evolution for the conclusion of the Goldstone theorem is emphasized.
\end{abstract}


\mbox{}

\relax \@setckpt{pref0}{ \setcounter{page}{1}
\setcounter{equation}{0} \setcounter{enumi}{0}
\setcounter{enumii}{0} \setcounter{enumiii}{0}
\setcounter{enumiv}{0} \setcounter{footnote}{0}
\setcounter{mpfootnote}{0} \setcounter{part}{0}
\setcounter{chapter}{-1} \setcounter{section}{0}
\setcounter{subsection}{0} \setcounter{subsubsection}{0}
\setcounter{paragraph}{0} \setcounter{subparagraph}{0}
\setcounter{figure}{0} \setcounter{table}{0}
\setcounter{Theorem}{0} }

\makeatletter
\noindent 1. {\em Infinitely extended quantum systems}

In the development of theoretical physics, the standard way of describing a broken symmetry has 
been that of introducing an explicit non-symmetric term in the equations of motion. A real 
revolution occurred with the realization of a much more economical and powerful mechanism, called 
{\bf spontaneous symmetry breaking}, by which symmetry breaking may be realized even if the 
equations of motion are symmetric (Heisenberg 1928, Nambu 1960, Goldstone 1961).
    

The mechanism of spontaneous symmetry breaking for classical systems has a counterpart for 
quantum systems. For such a transcription, in analogy with the classical case, a crucial role is 
played by the {\em infinite extension} of the system and {\em locality}, so that each ground state defines a physically disjoint realization of the system (i.e. a disjoint physical world or phase). For this purpose, it is useful to recall that infinitely extended quantum systems are conveniently described by {\em 
algebras $\A$ of local operators} (Haag and Kastler 1964).

For the sake of concreteness, for infinitely extended non-relativistic systems which allow a 
description in terms of canonical variables, one can take the {\em algebra $\A$ of localized 
canonical variables} generated by $\psi(f) = \int d^3 x \,\psi(\x) \bar{f}(\x)$, where $\psi(\x)$ 
is the canonical field ``localized`` in $\x$, $$[\,\psi(\x), \psi(\y)\,] = 0, 
\,\,\,\,\,\,[\psi(\x), \,\psi^*(\y)\,] = \delta(\x- \y),$$ and $f \in C^\infty(\Rbf^3)$, of 
compact support (i.e. $\in \D(\Rbf^3)$) or of fast decrease (i.e. $\in \S(\Rbf^3)$). $\psi(\x)$ 
and $\psi^*(\x)$ have the meaning of destruction and creation operators and in the Fock 
representation they respectively destroy and create elementary excitations localized at $\x$. For 
relativistic quantum fields, one can take the {\em local field algebra $\F$} generated by the 
local fields $\psi(f) = \int d^4 x \,\psi(x)\,f(x)$, $f \in \S(\Rbf^4)$. In both cases, space and 
time translations are assumed to be well defined transformations on the local algebras.

 In the case of a finite number $2N$ of degrees of freedom, described by the canonical variables 
$q_k, p_k$, $k = 1, ...N$, (hereafter referred to as the finite dimensional case), under general 
regularity conditions the Stone-von Neumann theorem (Stone 1930, von-Neumann 1931) states that 
the Schroedinger representation is the unique irreducible representation of the algebra $\A_c$ of 
canonical variables, up to unitary equivalence (see below).

On the other hand, the local algebras $\A$ which describe infinitely extended systems have many 
inequivalent representations and the first step is the selection of those which are physically 
relevant. We recall that a representation $\pi(\A)$ of $\A$ is a homomorphism of $\A$ (i.e. a 
mapping which preserves all the algebraic relations) into an algebra of operators in a Hilbert 
space $\H_\pi$, (with the technical condition of a common invariant dense domain).

Strong physical considerations motivate the following general conditions (in the case of zero 
temperature considered below). Such conditions apply to both non-relativistic and relativistic 
systems; for the latter ones, further conditions are needed corresponding to relativistic 
invariance and microscopic causality or (relativistic) locality (see below). 

In the following, 
when we are dealing with a given representation $\pi$, in order to simplify the notation we shall 
often denote by $A$ the representative $\pi(A)$ of the element $A$.

\vspace{1mm}\noindent {\bf I}. ({\bf Existence of energy and momentum}) In the representation 
$\pi$ of $\A$ in terms of operators in a Hilbert space $\H_\pi$, the space and time translations, 
$\a_\x, \,\a_t$, are described by strongly continuous groups of unitary operators $U(\x)$, 
$U(t)$, $\x \in \Rbf^s$, $t \in \Rbf$: $$\a_\x(A) = U(\x) A U(\x)^{-1}, \,\,\,\,\,\a_t(A) = U(t) 
A U(t)^{-1}, \,\,\,\forall A \in \A.$$

The strong continuity is equivalent to the existence of the generators, namely of the momentum 
${\bf P}$ and of the energy $H$.

\noindent {\bf II}. ({\bf Stability or spectral condition}) The energy spectrum $\sigma(H)$ is 
bounded from below. The relativistic invariant form of such a condition is $\sigma(H) \geq 0$, 
$H^2 - {\bf P}^2 \geq 0$.

Clearly, the failure of the spectral condition implies that the system may undergo a decay into 
states of lower and lower energy, i.e. a collapse.

\noindent {\bf III}. ({\bf Ground state}) Inf\,$\sigma(H)$ is a proper (non-degenerate) 
eigenvalue of $H$, the corresponding eigenvector $\Psio$ is called the ground state and it is a 
cyclic vector for $\pi(\A)$, i.e. $\pi(\A)\,\Psio$ is the common {\em dense} domain. $\Psio$ is 
also assumed to be invariant under space translations (or at least under a subgroup of them, but 
for simplicity, in the following, we shall restrict our discussion to the case of invariance 
under the full translation group)

\noindent {\bf IV}. ({\bf Local structure}) {\em Asymptotic abelianess}: for any two elements $A, 
B \in \pi(\A)$ $$weak-\lim_{|\x| \ra \infty} \,[\,\a_\x(A), \, B \,] = 0,$$ (the weak limit means 
the limit of the matrix elements for any pair of vectors belonging to $\H_\pi$)

If $A$ and $B$ describe quantities which can be measured, briefly if they are {\em observable} 
elements, asymptotic abelianess means that the measurement of $B$ becomes compatible with the 
measurement of $A$ (in the quantum mechanical sense), when the latter is translated at infinite 
(space) distance.

\def \lan {\langle} \def \ran {\rangle} Equivalently, the local structure may be codified by the 
{\em cluster property}: $$\lim_{|\x| \ra \infty} [ \langle A \,\a_\x(B) \rangle - \langle A 
\rangle\,\langle B \ran ] = 0,\,\,\,\,\,\forall A, \,B \in \pi(\A), $$ where the bracket $\lan 
\,\ran$ denotes the ground state expectation, i.e. $\lan A \ran \eqq (\Psio, A \Psio)$.

The physical meaning of the cluster property is that a decorrelation occurs for infinitely 
separated elements of the local algebra, as it must be for an acceptable physical interpretation. 
A very important result is that the validity of the cluster property is equivalent to the 
uniqueness of the translationally invariant state in $\H_\pi$. For brevity, a representation with 
a unique translationally invariant ground state shall be called a {\em pure phase}.

\vspace{1mm} In conclusion, an infinitely extended quantum system is conveniently described by 
the local algebra $\A$ and by the set of its physically relevant representations; we denote by 
$\Sigma$ the set of the corresponding states.

 Inequivalent representations of $\A$ describe disjoint realizations of the system and the 
possibility arises of the mechanism of symmetry breaking, as in the classical case.

\vspace{2mm}

\noindent 2. {\em Symmetries of a quantum system and their spontaneous breaking}
\vspace{1mm}

In the case of finite dimensional quantum systems, described by the Schroedinger representation, 
as clarified by Wigner (Wigner 1959), a {\em symmetry} $g$ is an invertible mapping of the 
states, i.e. of the rays, of the Schroedinger Hilbert space $\H_S$ and, as a consequence of 
Wigner theorem (Wigner 1959), it can be implemented by a unitary operator $U(g)$ in $\H_S $. 
Hence, $g$ defines an algebraic mapping $\a_g$ of the algebra of canonical variables $\A_c$ 
(since $\A_c$ is faithfully represented in $\H_S$): $$\a_g(A) = U(g)\,A \,U(g)^{-1}, 
\,\,\,\,\forall A \in \A_c, $$ which preserves all the algebraic relations (technically $\a_g$ 
defines an automorphism of $\A_c$), i.e. an {\em algebraic symmetry}.

The relevant point is that, in the case of infinitely extended systems there may be automorphisms 
$\a_g$ of the local algebra $\A$ which are not described by unitary operators in one 
representation $\pi $ of $\A$. This means that $g$ exists as a symmetry at the algebraic level, 
but it is not a symmetry of the realization of the system provided by the representation $\pi$. 
In this case, the {\bf symmetry $g$ is broken} in $\H_\pi$. As in the classical case, this very deep mechanism cannot be reduced to the statement that "spontaneous symmetry breaking occurs in a situation where, given a symmetry of the equation of motion, solutions exists which are not invariant under the action of this symmetry without any explicit asymmetric input" (Stanford Encyclopedia of Philosophy 2008). 
The {\bf symmetry} is {\bf unbroken} 
in the realization of the system described by a representation $\pi$ if it is implemented by 
unitary operators in $\H_\pi$.

In the following, we consider the case of {\em internal symmetries}, i.e. symmetries which 
commute with space and time translations. For them one has the following criterion of symmetry 
breaking, which is at the basis of the standard discussion.

\begin{Theorem} Given an internal symmetry $\b$, i.e. an automorphism of $\A$, which commutes 
with space and time translations, a necessary and sufficient condition for $\b$ being broken in a 
representation $\pi$, satisfying conditions I-IV, is that there exists $A \in \pi(\A)$ such that 
$\lan \b(A) \ran \neq \lan A \ran$. In this case $\lan A \ran $ is called a {\bf symmetry 
breaking order parameter}. \end{Theorem} It is worthwhile to remark that the validity of the 
cluster property, i.e. the uniqueness of the translationally invariant ground state in $\H_\pi$, 
is crucial for the effectiveness of the criterium. In fact, the cluster property is not satisfied 
in a representation defined by a mixed state and the invariance of the expectations of a mixed 
ground state does not exclude that the symmetry is broken in each pure phase in which the 
representation decomposes.

\vspace{2mm} \noindent 3. {\em Goldstone theorem without assuming relativistic invariance}
\vspace{1mm}

A very important consequence of the breaking of a continuous one-parameter Lie group of 
symmetries $\b^\l$, $ \lambda \in \Rbf$, in a representation $\H_\pi$, is the {\bf Goldstone 
theorem} (Goldstone 1961, see also Nambu 1960). The theorem provides {\em exact information on 
the energy momentum spectrum} of the intermediate states which saturate the two point function 
$\lan j_\mu(x)\,A \ran$, where $A$ defines the symmetry breaking order parameter and $j_\mu$ is 
the conserved current whose existence follows from the symmetry of the dynamics (typically from 
the invariance of the Lagrangian under $\b^\l$), hereafter referred to as the {\em conserved 
Noether current}.

The formulation and proof in the non-relativistic case has been the subject of debate, in our 
opinion because it has not been clearly realized that the crucial hypothesis is a sufficient 
locality of the dynamics. Since there are interesting physical systems which exhibit the breaking 
of a continuous symmetry with an apparent evasion of the conclusions of the Goldstone theorem 
(like superconductivity, Coulomb systems and plasmons, Higgs mechanism etc.), it is worthwhile to 
examine the hypotheses with special care, even at the risk of looking pedantic.

The original argument by Goldstone (Goldstone 1961) relied on a semi-classical approximation 
based on a mean field ansatz. The substantial improvement leading to a proof of the theorem by 
Goldstone, Salam and Weinberg (GSW) (Goldstone, Salam, Weinberg 1961), was the realization and 
the exploitation of the link between the generation of the one-parameter symmetry group $\b^\l$ 
by the algebraic derivation on $\A$: $$ (d/d \l) \lan \b^\l(A) \ran |_{\l=0} \eqq \lan \delta A 
\ran $$ and the integral of the charge density $j_0$ of the conserved Noether current $j_\mu $, 
$\partial^\mu j_\mu = 0$.
   
Formally, such a link reads \be{ \lan \delta A \ran = \limR \lan [\,Q_R(t), \,A\,] \ran = \limR \lan 
[\,Q_R(0), \,A\,] \ran, }\ee $$ Q_R(t) \eqq \int_{|\x| \leq R} d^3 x\,j_0(\x, t).$$
  The above eq.\,(0.1), which is at the basis of the GSW proof for relativistic systems (to be 
discussed below), is also the {\em crucial} ingredient of the Goldstone theorem for 
non-relativistic systems. In the latter case, a proper formulation and discussion of eq.\,(0.1), 
which shall be later referred to as the property of {\bf local generation of $\b^\lambda$ by the 
conserved Noether current $j_\mu$}, requires a special care, mainly because the dynamics is not 
strictly local as in the relativistic case (see below). The point is that the check of the 
generation of the symmetry by the current density at equal times (which requires only the use of 
the canonical commutation relations) is not enough for establishing eq.\,(0.1). The statement (which may be found in the literature) that the independence of time of the commutator $[ Q(t), A ]$ is guaranteed by the invariance of the Hamiltonian, since $ \dot{Q} = i [ H, Q ] = 0$, is not correct. For the local 
generation at unequal times,  the local property of the dynamics plays a crucial role. In 
fact, if the dynamics is strictly local, i.e. the time evolution of an operator localized in a 
compact set is still localized in a (possibly larger) compact set, the generation at equal times 
implies eq.\,(0.1). However, such an implication fails if the dynamics induces a long range 
delocalization (see below) (Morchio and Strocchi 1985, Strocchi 2008) and the Goldstone theorem does not apply.

The first step in the discussion of eq.\,(0.1) is the condition of integrability of the charge 
density commutators. Technically, one must have that $J(\x, t) \eqq i \lan [\,j_0(\x, t), \,A\,] 
\ran$ is absolutely integrable in $\x$ as a tempered distribution in $t$, i.e. after smearing 
with any test function $h(t)$ of fast decrease ({\bf charge integrability condition}). This 
condition is usually overlooked in the standard treatments of the non-relativistic Goldstone 
theorem, but it is needed for the continuity of the energy momentum spectrum in the neighborhood 
of $\k = 0$, which is necessary for drawing the conclusions of the theorem (Swieca 1967, Morchio 
and Strocchi 1985, 1987, Strocchi 2008).

The charge integrability condition is satisfied if the dynamics is sufficiently local, so that 
the (unequal time) commutators decay sufficiently fast in the limit of infinite space 
separations; the condition is obviously satisfied in the case of a strictly local dynamics, as it 
occurs in relativistic theories described by fields satisfying micro-causality.

The localization of the dynamics plays a crucial role also for assuring the time independence of 
$\limR \lan [\,Q_R(t), \,A\,] \ran$, a necessary condition for its equality to the expectation of 
the time independent derivation $ \delta A$.  
(anti)commutation relations are local, the issue is the effect of 
usually induced by the time evolution.

Since, the continuity equation gives $$(d/ d t) \lan [\,j_0(\x, t),\, A\,] \ran = \mbox{div} \lan 
[\,\j(\x, t), A\,] \ran, $$ the time independence of $\limR \lan [\,Q_R(t), \,A\,] \ran $ is 
guaranteed if the time evolution is sufficiently local, so that the {\bf Swieca condition} 
(Swieca 1967) is satisfied ($s$ denotes the space dimension) $$\lim_{|\abf| \ra \infty} 
\,|\abf|^{s - 1}\,\lan [\,\j(\x+\abf, 0), \,\a_t(A)\,] \ran = 0.$$ Such a condition and more 
generally the needed time independence of $\limR \lan [\,Q_R(t), \,A\,] \ran$, which may actually 
require a weaker form of the Swieca condition (Strocchi 2005), is violated in the case of 
interactions described by a two body potential decaying as $1/r$ and this is the explanation of 
the evasion of the Goldstone theorem by Coulomb systems (as well as in the closely related Higgs 
mechanism in the Coulomb gauge). \goodbreak

\vspace{1mm} In fact, roughly the logic of the theorem is the following: the time independence of 
$$\int d x J(\x, t) = \lim_{\k \ra 0} \,\int d \omega \,e^{i \omega t}\, \tilde{J}(\k, \omega), 
\,\,\,\,J(\x, t) \eqq i \lan [\,j_0(\x, t), \,A\,] \ran,$$ implies that $\lim_{\k \ra 
0}\,\tilde{J}(\k, \omega) = \l \delta(\omega)$, ($\l$ a suitable constant).

If there is symmetry breaking {\em and} $\b^\l$ is locally generated by $j_\mu$, one has $\int d 
x J(\x, t) = \lan \delta A \ran \neq 0$, which implies $\l \neq 0$. Thus, there cannot be a 
positive constant $\mu$, such that the energy spectrum of the intermediate states, which saturate 
the two point function $J(\x, t)$, is supported in $\{\omega(\k) \geq \mu > 0\}$, in the 
neighborhood of $\k = 0$; hence, there cannot be an energy gap $\mu$.

As a consequence of the above discussion one has the following (Lange 1966; Morchio and Strocchi 
1987) \begin{Theorem} ({\bf Goldstone theorem for non-relativistic systems}) If \newline i) 
$\b^\l, \l \in \Rbf$ is a one-parameter internal symmetry group, \newline ii) $\b^\l$ is locally 
generated by the conserved current $j_\mu$ (which transforms covariantly under space and time 
translations), eq.\,(0.1), \newline iii) $\b^\l$ is broken in the representation $\pi$ with 
translationally invariant ground state vector $\Psio$, i.e. there is $A \in \A$ such that $\lan 
\delta A \ran \neq 0$,

\noindent then there are quasi particle excitations with infinite lifetime in the limit $\k \ra 
0$, i.e. particle-like excitations with energy width $\Gamma(\k) \ra 0$ as $\k \ra 0$, and with 
energy $\omega(\k) \ra 0$ as $\k \ra 0$ ({\bf Goldstone quasi particles}); the corresponding 
states have non-trivial projections in the subspaces $\{ \pi(\a_t(A)) \Psio, \,t \in \Rbf\}$, $\{ 
\pi(Q_R) \Psio,\, R \in \Rbf\}$. \end{Theorem}

 \vspace{2mm} \noindent 4. {\em Goldstone theorem for relativistic quantum systems}
\vspace{1mm}

For relativistic quantum systems described by a {\em local} field algebra $\F$, the relevant 
representations are selected by the following physical conditions which strengthen the conditions 
I-IV discussed above, by incorporating relativistic invariance and relativistic locality: 
\vspace{1mm} \newline {\bf I.} ({\bf Poincar\'{e} covariance}) The automorphisms $\a(a, 
\Lambda)$, ($ a \in \Rbf^4$, $\Lambda$ denoting the elements of the restricted Lorentz group), 
which describe the transformations of the Poincar\'{e} group are implemented by a strongly 
continuous group of unitary operators $U(a, \Lambda(M))$, $ M \in SL(2, C)$,

\noindent {\bf II.} ({\bf Relativistic spectral condition}) $ H \geq 0, \,\,\,\,\,H^2 - {\bf P}^2 
\geq 0,$

\noindent {\bf III.} ({\bf Vacuum state}) There is a unique space-time translationally invariant 
state $\Psio$ (vacuum state) cyclic for the field algebra $\F$

\noindent {\bf IV.} ({\bf Locality}) The algebra $\F$ satisfies relativistic locality, i.e. fields commute or anticommute at relatively spacelike points.

\vspace{2mm}

In the case of relativistic local fields, the inevitable ultraviolet singularities require a more 
careful regularization of the integral of the charge density $j_0$, e.g. $$Q_R = j_0(f_R, h) = 
\int d^4 x\,f_R(\x)\,h(x_0)\,j_0(\x, x_0),$$ where $f_R(\x) = f(|\x|/R)$, $f \in \D(\Rbf)$, $f(y) 
= 1,$ if $|y| < 1$, $f(y) = 0$, if $|y| > 1 + \eps$, $h \in \D(\Rbf),$ supp\,$h \subset [ - 
\delta, \,\delta ]$, $\tilde{h}(0) = \int d x_0\,h(x_0) = 1.$

\vspace{1mm}In the case of relativistic local fields the problems discussed above for 
non-relativistic systems, like the charge integrability condition, the local generation of 
$\b^\l$ and the time independence of the charge density commutators do no longer arise thanks to 
locality. Then, one has (Goldstone, Salam and Weinberg 1962; Kastler,Robisson and Swieca 1966)

\begin{Theorem} ({\bf Goldstone theorem for relativistic local fields}) Let
 $\b^\l$, $\l \in \Rbf$, be a one parameter group of internal symmetries locally generated by the 
covariant conserved current $j_\mu$, i.e. $\forall F \in \F$ $$ \lan \delta F \ran = \limR \,i 
\lan [\,Q_R,\, F\,] \ran;$$ then, the symmetry breaking condition $\lan \delta F \ran \neq 0$ 
implies that the Fourier transform $\tilde{J}_\mu(k)$ of the two point function $\lan j_\mu(x)\,F 
\ran$ contains a $ \delta(k^2)$ contribution, i.e. there are {\bf Goldstone massless modes}.  
\end{Theorem}
 
The proof of the theorem is particularly simple if $F$ is a scalar elementary field $\ph(x)$, 
i.e. a field transforming as a pointlike operator under the Poincar\'{e} group: $$U(a, \Lambda) 
\,\ph(x)\,U(a, \Lambda)^{-1} = \ph(\Lambda x + a).$$ This was indeed the case considered in the 
original proof by Goldstone, Salam and Weinberg. In fact, the Poincar\'{e} covariance of the two 
point function $J_\mu(x, y) \eqq \lan j_\mu(x)\,\ph(y) \ran$ implies that it has the following 
form $$J_\mu(x, y) = J_\mu(x-y) = \partial_\mu J(x - y),$$ with $J(x)$ a Lorentz invariant 
function. Then, the current conservation gives $$ 0 = \partial^\mu J_\mu(x) = \square J(x).$$ 
This means that $\tilde{J}_\mu(k) = \lambda k_\mu \delta(k^2)$, with $\lambda \neq 0$ as a 
consequence of the symmetry breaking condition $\lan \delta \ph \ran \neq 0$; hence, the 
intermediate states which saturate the two point function $J_\mu(x)$ have energy-momentum 
spectrum supported in $k^2 = 0$, i.e. they describe massless modes.

\vspace{1mm} The role of Lorentz covariance in the original proof (Goldstone, Salam and Weinberg 
1962) has led to the belief that its failure provides the relevant mechanism for evading the 
conclusions of the theorem. This was indeed emphasized by Higgs (Higgs 1964) as the property of 
the Coulomb gauge which allowed the evasion of the Goldstone theorem for the Higgs mechanism. As 
stressed in the non-relativistic case, the crucial property is rather the localization property 
of the dynamics. As a matter of fact, the more general proof of the theorem, given by Kastler, 
Robinson and Swieca (Kastler, Robinson, Swieca 1967), covers the case of a symmetry breaking 
order parameter $F$ without a pointlike structure, like a compound field or a polynomial of 
elementary fields, so that the two point function $\lan j_\mu(x)\,F \ran$ does not have simple 
transformation properties under Lorentz boosts.

In view of the crucial role of locality, a few comments may be relevant about its validity in the 
case of relativistic systems. Clearly, the algebra of observable fields and in particular the 
algebra $\A_{obs}$ of their bounded functions, usually called the {\em algebra of observables}, 
must satisfy locality, since measurements of operators localized in spacelike separated regions 
must be independent in the quantum mechanical sense.

However, this does not imply that the states of the system, operationally defined by their 
expectations of $\A_{obs}$, are {\em local states}, i.e.  can be obtained by applying local 
fields to a vacuum state. Thus, one may have to use a non-local field algebra to reach such 
non-local states. In fact, this is the case of the charged states in quantum electrodynamics, 
whose description requires the non-local charged fields of the Coulomb gauge. Then, if the 
symmetry breaking is realized by the vacuum expectation of a non-local field, the same problems 
of the non-relativistic systems arise and the {\em same mechanism} of evasion of the Goldstone 
theorem may take place.

\vspace{3mm} \noindent 5. {\em Appendix}

In the seminal paper of 1964, Haag and Kastler emphasized the role of locality and, on the basis 
of mathematical and physical considerations, argued that the convenient mathematical setting for 
the description of infinitely extended systems is that of {\em $C^*$-algebras}; such algebras are 
(isomorphic to) norm closed algebras of operators in a Hilbert space, stable under the adjoing 
operation (Gelfand and Naimark 1943).

The abstract definition of a {\bf $C^*$-algebra} $\A$ is that it is a linear associative algebra 
over the field $\Cbf$ of complex numbers, with an involution $^*$ and equipped with a norm $|| 
\,||$ satisfying

\noindent i) $|| A B || \leq ||A||\,||B||$

\noindent ii) $||A^*\,A|| = ||A||^2$

\noindent iii) $\A$ is complete with respect to the norm.

In the following, we shall always consider {\em unital} $C^*$-algebras $\A$, i.e.  with an 
identity $\id$, ($\id A = A = A \id$, $\forall A \in \A$).

A {\bf representation} $\pi$ of a $C^*$-algebra $\A$ in a Hilbert space $\H_\pi$ is a 
homomorphism of $\A$ into the $C^*$-algebra $\B(\H_\pi)$ of the bounded operators in $\H_\pi$, 
namely a mapping which preserves all the algebraic relations, including the $^*$ (i.e. $\pi$ is 
linear, multiplicative $\pi(A B) = \pi(A)\,\pi(B)$ and $\pi(A)^* = \pi(A^*))$. The property of 
preserving the $^*$ is spelled out by speaking of $^*$-homomorphism, but for simplicity we shall 
omit the $^*$. A homomorphism of $\A$ into the $C^*$-algebra $\B$ which is one-to-one and onto 
(bijective) is an {\em isomorphism}; an isomorphism of $\A$ onto itself is called an {\em 
automorphism}. A vector $\Psi \in \H_\pi$ is {\em cyclic} if $\pi(\A)\, \Psi$ is dense in 
$\H_\pi$. A representation $\pi$ is {\em faithful} if $ker \pi = \{0\}$, i.e. $\pi(A) = 0$ 
implies $A = 0$.

A {\em state} $\omega$ on $\A$ is a positive linear functional on $\A$, i.e. $\forall A, B \in 
\A$ $$\omega(\lambda A + \mu B) = \lambda \,\omega(\A) + \mu \,\omega(B), \,\,\,\lambda, \mu \in 
\Cbf,\,\,\,\,\,\omega(A^*\,A) \geq 0.$$ A state $\omega$ on $\A$ defines a representation 
$\pi_\omega(\A)$ of $\A$ in a Hilbert space $\H_\omega$, i.e. a homomorphism of the $C^*$-algebra 
$\A$ into the $C^*$ algebra of bounded operators in $\H_\omega$. Moreover, $\H_\omega$ contains a 
cyclic vector $\Psi_\omega$ such that $\omega(A) = (\Psi_\omega, \,\pi_\omega(A)\,\Psi_\omega)$. 
The representation is called the Gelfand-Naimark-Segal (GNS) representation defined by $\omega$ 
and it is unique up to isomorphisms (Gelfand and Naimark 1943, Segal 1947).

A state $\omega$ on $\A$ is called {\em pure} if it cannot be written as a convex linear 
combination of other states, and {\em mixed} otherwise.  A mixed state $\omega$ on $\A$ cannot be 
described by a state vector in an irreducible representation of $\A$; in fact, it is described by 
a density matrix: $$\omega(A) = \mbox{Tr}\, (\rho_\omega \pi(A)), \,\,\, \forall A \in \A,$$ $$ 
\rho = \sum_i \l_i P_i, \,\,\, \l_i \geq 0, \,\,\,\sum_i \l_i = 1, $$ where $P_i$ are 
one-dimensional orthogonal projections.

Clearly, any vector $\Psi \in \H$, with $\H$ the carrier of a representation $\pi$ of the 
$C^*$-algebra $\A$, defines a state $\omega_\Psi$ on $\A$, $\omega_\Psi(A) \eqq (\Psi, 
\,\pi(A)\,\Psi)$ and the corresponding GNS representation is unitarily equivalent to $\pi$.

For quantum systems with a finite number $2 N$ of degrees of freedom, described by the canonical 
variables $q_k, p_k$, $ k = 1, ...N$, the standard $C^*$-algebra is the canonical Weyl algebra 
$\A_W$, generated by the unitary operators $U(\alpha)$, $V(\beta)$, $\alpha, \beta \in \Rbf^N$, 
formally given by the exponentials $U(\a)= e^{i \sum_k \a_k q_k}$, $V(\b) = e^{i \sum_k \b_k 
p_k}$, with the commutation relations induced by the canonical commutation relations of the 
canonical variables $q_k, p_k$. In particular, $U(\a)$, $V(\b)$ define $N$-parameter abelian 
unitary groups. A representation $\pi$ of the Weyl algebra is {\em regular} if $\pi(U(\a))$, 
$\pi(V(\b))$ are weakly continuous in $\a, \b$, i.e. their matrix elements are continuous 
functions of $\a, \,\b$; regularity is a necessary and sufficient condition for the existence of 
the generators, i.e. of the canonical variables $q_k, p_k$ (as unbounded operators in $\H_\pi$).  
The Stone-von Neumann theorem (Stone 1930, von Neumann 1931) states that, up to unitary 
equivalence, there is only one regular irreducible representation of the canonical Weyl algebra. 
This means that there is only one (regular) realization of a system described by a canonical Weyl 
algebra. This implies that, for finite dimensional quantum systems, symmetries defined on the canonical variables $q, p$ can never be broken (Strocchi 2008); the standard argument that symmetry breaking is prevented by tunneling (Stanford Encyclopedia of Philosophy 2008) looks less general and in fact does not apply to finite dimensional spin systems, for which an analog of the Stone-von Neumann uniqueness theorem may be proved.

 As advocated by Haag and Kastler, infinitely extended systems should be described by local 
$C^*$-algebras, i.e. by norm closed $^*$-algebras generated by operators which are localized.

In the non-relativistic case, one may consider the canonical field operators $\psi(\x)$, 
$\psi^*(\x)$, localized at $\x $, smear them with test functions $f, g \in C^\infty(\Rbf^3)$, of 
compact support (i.e. $\in \D(\Rbf^3)$) or of fast decrease (i.e. $\in \S(\Rbf^3)$), and take the 
$C^*$-algebra $\A$ generated by the exponentials $$U(f) = e^{i(\psi(f) + \psi^*(f))},\,\,\,V(g) = 
e^{ \psi(g) - \psi^*(g)}.$$ This algebra represents the infinite dimensional analog of the Weyl 
algebra; the Stone-von Neumann theorem does not apply and in fact there are many inequivalent 
representations.

In the case of relativistic systems, Haag and Kastler proposed to define a quantum field theory 
by the association of a $C^*$-algebra $\A(\O)$ to each bounded region $\O$, typically a double 
cone, in Minkowski space, with the following setting: i) {\em isotony}: if $\O_1 \supset \O_2$, 
then $\A(\O_1) \subset \A(\O_2)$; ii) {\em locality}: if $\O_1$ and $\O_2$ are completely 
spacelike with respect to each other, then $\A(\O_1)$ and $\A(O_2)$ commute; iii) the norm 
completion $\A$ of the set theoretical union of all $\A(\O)$, called the {\bf algebra of 
observables}, contains all the observables; iv) {\em relativistic covariance}: the inhomogeneous 
Lorentz group is represented by automorphisms $A \ra \a_L(A)$ such that $\a_L(\A(\O)) = 
\A(\O_L)$, $ \forall A \in \A$, with $O_L$ the image of $\O$ under the inhomogeneous Lorentz 
transformation $L$.

The algebra $\A(\O)$ has the meaning of the observables which can be measured in the space-time 
region $\O$. The $C^*$ structure emphasizes that fact that, from an operational point of view, 
observables should be described by bounded operator (Segal 1947, Strocchi 2005). From a 
mathematical point of view, there is a big advantage in using bounded operators, since e.g. 
domain questions do not arise. For simplicity, we shall not spell out the distinction between the 
union of the $\A(\O)$, called the local observable algebra, and its completion called the 
quasilocal algebra of observables. Furthermore, it is taken for granted that $\A$ admits a 
faithful algebraically irreducible representation.

The standard formulation of quantum field theory makes use of local quantum fields, which are 
unbounded operators and therefore cannot be elements of a $C^*$-algebra. Furthermore, there are 
quantum fields which do not describe observable quantities; e.g. a fermion field is not an 
observable field, whereas so is the electromagnetic field $\Fmn$. Thus, the algebra $\F$ 
generated by the local fields contains the subalgebra $\F_{obs}$ of observables fields; the 
$C^*$-algebra of observables should be obtained by considering bounded functions of the 
observable fields. For example, $e^{i \Fmn(f)}$, with $f \in \D(\Rbf^4)$, is a bounded operator 
and may be considered as an element of the $C^*$-algebra of observables.
  
In the Haag-Kastler approach, called the {\em algebraic} approach to {\em quantum field theory}, 
a symmetry is an automorphism of the algebra of observables and it is broken in a representation 
$\pi$ if it is not implementable by a unitary operator there.

\vspace{2mm}
\noindent Acknowledgments. I am  indebted to Riccardo Guida for useful  comments in the preparation of this note.

\vspace{5mm}\newpage REFERENCES

\noindent P.W. Anderson, Phys. Rev. {\bf 130}, 439 (1963)

\noindent T. Eguchi and K. Nishijima, {\em Broken Symmetry. Selected papers by Y. Nambu}, World 
Scientific 1995

\noindent I. Gelfand and M.A. Naimark, On the imbedding of normed rings into the ring of 
operators in a Hilbert space, Mat. Sborn., N.S., {\bf 12}, 197-217 (1943)

\noindent J. Goldstone, Field Theories with Superconductor Solutions, Nuovo Cim. {\bf 19}, 
154-164 (1961)

\noindent J. Goldstone, A. Salam and S. Weinberg, Broken Symmetries, Phys. Rev. {\bf 127}, 
965-970 (1962)

\noindent R. Haag, {\em Local Quantum Physics}, Springer 1996

\noindent R. Haag and D. Kastler, Algebraic approach to quantum field theory, Jour. Math. Phys. 
{\bf 5}, 848-861 (1964)

\noindent W. Heisenberg, Zut Theorie des Ferromagnetism, Z. Physik {\bf 49}, 619-636 (1928)

\noindent P.W. Higgs, Broken Symmetries, Massless Particles and Gauge Fields, Phys. Lett. {\bf 
12}, 132-133 (1964)

\noindent D. Kastler, D.W. Robinson and J.A. Swieca, Conserved currents and associated 
symmetries; Goldstone's theorem, Comm. Math. Phys. {\bf2}, 108-120 (1962)

\noindent R.V. Lange, Nonrelativistic Theorem Analogous to the Goldstone Theorem, Phys. Rev. {\bf 
146}, 301-303 (1966)

\noindent G. Morchio and F. Strocchi, Infrared problem, Higgs phenomenon and long range 
interactions, Lectures at the Erice School of Mathematical Physics, Erice 1985, {\em Fundamental 
Problems of Gauge Field Theories}, G. Velo and A.S. Wightman eds. Plenum 1986

\noindent G. Morchio and F. Strocchi, Mathematical structures for long-range dynamics and 
symmetry breaking, Jour. Math. Phys. {\bf 28}, 622-735 (1987)

\noindent Y. Nambu, Axial vector current conservation in weak interactions, Phys. Rev. Lett. {\bf 
4}. 380-382 (1960)

\noindent Y. Nambu, Quasiparticles and Gauge Invariance in the Theory of Superconductivity, Phys. 
Rev. {\bf 117}, 648-663 (1960)

\noindent I. Segal, Postulates of general quantum mechanics, Annals of Math., {\bf 48}, 930-948 
(1947)

\noindent Stanford Encyclopedia of Philosophy, Symmetry and Symmetry Breaking, 2008
 
\noindent M.H. Stone, Linear transformations in Hilbert space, III. Operational methods and group 
theory, Proc. Nat. Acad. Sci. U.S.A., {\bf 16}, 172-175 (1930)

\noindent F. Strocchi, {\em Symmetry Breaking}, Springer, 2005, 2nd ed. 2008

\noindent F. Strocchi, {\em An Introduction to the Mathematical Structure of Quantum Mechanics}, 
World Scientific 2005, 2nd enlarged edition 2010

\noindent J.A. Swieca, Range of forces and broken symmetries in many-body systems, Comm. Math. 
Phys. {\bf 4}, 1-7 (1967)

\noindent J. von Neumann, Die Eindeutigkeit der Schroedingerschen Operatoren, Math Ann. {\bf 
104}, 570-578 (1931)

\noindent E.P. Wigner, {\em Group Theory and Its Applications to the Quantum Mechanics of Atomic 
Spectra}, Academic Press 1959

\end{document}